\documentclass[summary]{ursi}

\usepackage{color,soul}
\usepackage{amsmath}

\title{Electromagnetic Illusion in Smart Environments}

\author{Hamidreza Taghvaee*\affref{ref1},
Mir~Lodro\affref{ref1},Neekar M Mohammed\affref{ref2}\affref{ref3},Sergio Terranova\affref{ref2},
Sendy Phang\affref{ref1}, Martin Richter\affref{ref2}, and Gabriele Gradoni\affref{ref1}\affref{ref2}}

\affiliation{%
  \aff{ref1}{George Green Institute for Electromagnetics Research, Department of Electrical and Electronics Engineering, University of Nottingham, Nottingham NG7 2RD, United Kingdom (e-mail: hamidreza.taghvaee@nottingham.ac.uk)}
  \aff{ref2}{School of Mathematical Sciences, University of Nottingham, Nottingham NG7 2RD, United Kingdom}
  \aff{ref3}{Department of Mathematics, College of Science, University of Sulaimani, Sulaymaniyah, 
Kurdistan Region, Iraq}
}

\begin{document}

\maketitle

\begin{abstract}
Metasurfaces can be designed to achieve prescribed functionality. Careful meta-atom design and arrangement achieve homogeneous and inhomogeneous layouts that can enable exceptional capabilities to manipulate incident waves. Inherently, the control of scattering waves is crucial in wireless communications and stealth technologies. Low-profile and light-weight coatings that offer comprehensive manipulation are highly desirable for applications including camouflaging, deceptive sensing, radar cognition control, and defense security. Here, we propose a method that achieves electromagnetic illusion without altering the object. A proof of principle is proposed and practiced for one-dimensional media. The idea is to engineer the environment instead of the object coating. This work paves the way for versatile designs that will improve electromagnetic security applications with the aid of smart environments.
\end{abstract}

\section{Introduction}
Any type of deceiving sense is an illusion. Naturally, there are different illusions associated with human senses. For instance, an optical illusion is a misleading visual grasp of a real object \cite{1}. A general extension of this concept is an electromagnetic (EM) illusion that confuses sensors or receiving antennas and thus has many potential protection-oriented applications. Since 2006, when Pendry et al \cite {2} and Leonhardt \cite{3} designed invisibility cloaks as an example of EM illusions, many kinds of illusion devices were designed and realized \cite{4,5}. For example, an illusion device can make an object look larger, appear at another place, or with different shapes and components \cite{6,7,8}.

Later, metasurfaces are introduced as a powerful platform due to their complete control over the scattered fields and advanced information encoding capabilities~\cite{9,10}. These features make them an excellent candidate for compact illusions devices. Recently, reconfigurable metasurfaces are used for dynamic illusions by controlling electronic elements~\cite{11}. A tunable graphene-based metasurface is designed by placing graphene ribbons on a dielectric cavity resonator. The wavefront of the field reflected from a triangular bump, covered by the metasurface activated by an electric bias, resembles the one generated by a spherical object~\cite{12}. 


There are a plethora of studies in EM illusions. However, all of them suggest adopting an object coating to form an illusion. In addition, these studies have been limited to free space, thus assuming no back-reflection from the environment. We argue that engineering the object itself may disrupt its function, and borrow complicated designs as well as manufacturing/installation procedures. Also, not in all the scenarios we have direct access to the object. For instance, to conceal the object from imaging/localization technologies, cloaking within a confined propagation environment, e.g., rooms, streets, should be achieved. The task of object illusion in complex environments is challenging and has attracted limited interest in the literature due to the complicated nature of the wave boundary value problems associated with it. Finding an appropriate mathematical description of the object-mirror-environment interaction process remains an open problem.

This scenario is radically different from free-space with regards to the physics of waves and mathematical modeling. These two aspects are intertwined and it is thus necessary to solve the boundary value-problem self-consistently accounting for environment, object and cloak altogether. The solutions present in the literature can thus be evolved to achieve object cloaking within a confined environment. From a mathematical point of view, the problem of object illusion in chaotic environments poses extraordinary challenges. It is a cascaded scattering process that can be modeled with products and sums of transfer matrices with arbitrary dimensions. 

\begin{figure}[htbp]
  \centering
  \includegraphics[trim={1cm 0 1cm 0},clip,width=82mm]{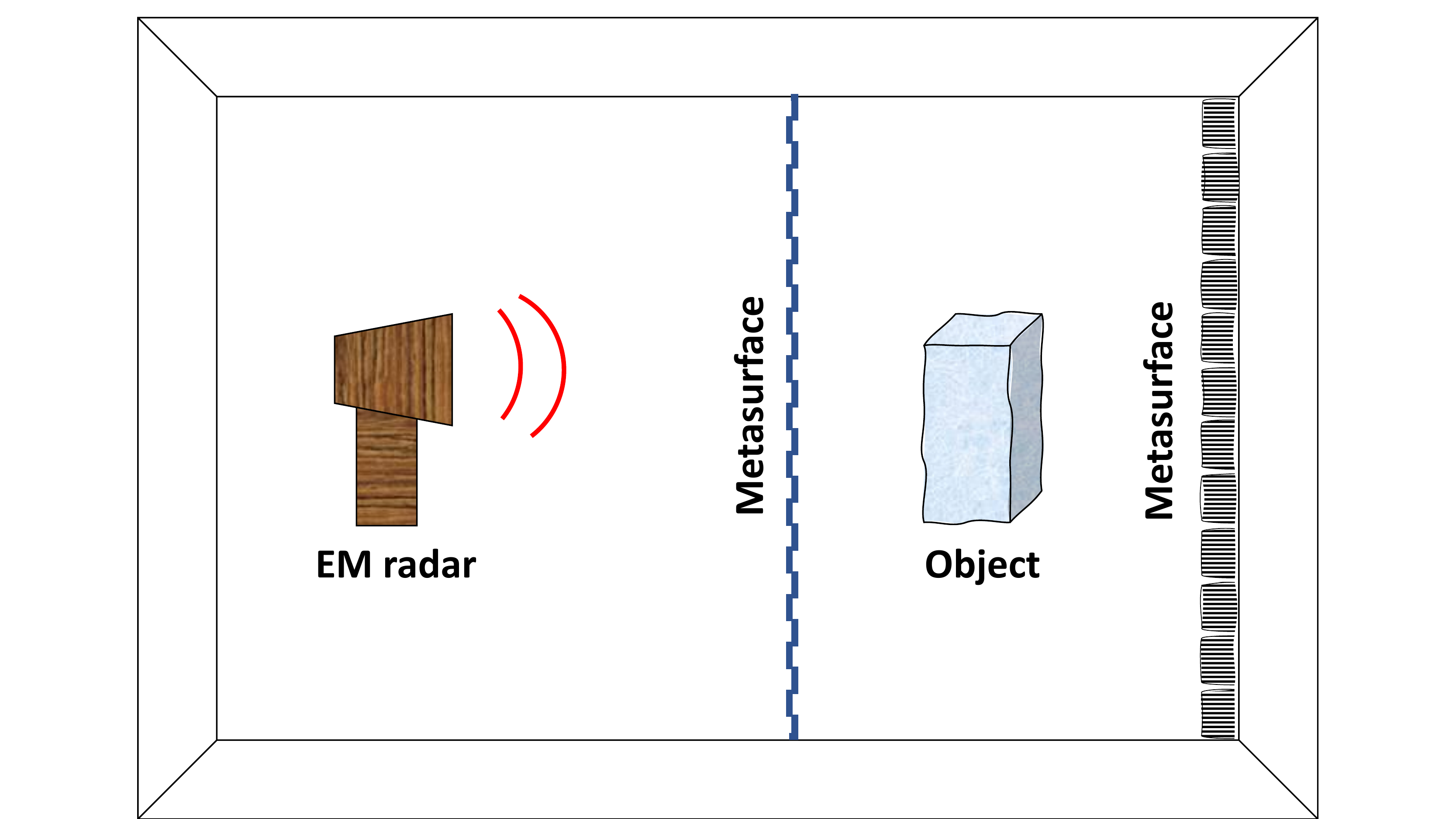}
  \caption{Illustration of the model under study in which electromagnetic illusion is created with a smart environment empowered by metasurfaces. Blue dashed lines represent a transmissive metasurface and black grooved lines represent a reflective metasurface.}
  \label{fig:ga}
\end{figure}

Here, we considered a 1D case of the problem as a proof of concept. Figure~\ref{fig:ga}, represents the cartoon of the model under study. The objective of this study is to engineer a transmissive metasurface before the object or a reflective metasurface behind the object that creates an EM illusion. The illusion consists of altering the object's scattering features to deceive the observer's sensing. In the following sections, we describe how to create the illusion by engineering the susceptibility of the transmissive metasurface or the surface impedance of the reflective metasurface. As an example, we consider the illusion of hiding the object. 

\section{Methodology}
\begin{figure}[htbp]
\vspace{-0.25cm}
  \centering
  \includegraphics[trim={2cm 2cm 2cm 3cm},clip,width=82mm]{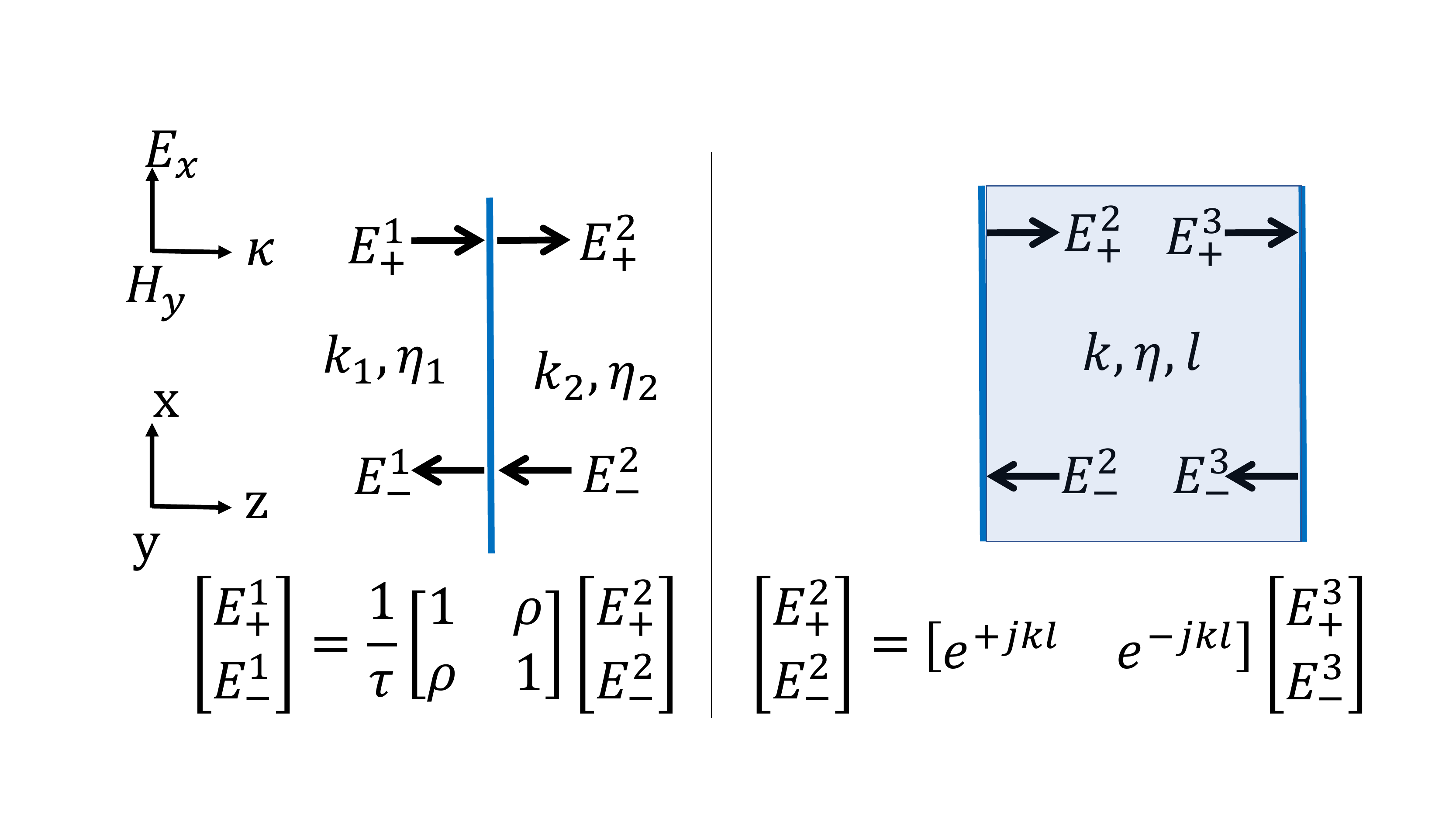}
  \caption{A continuous interface can be modeled as a $2\times2$ shift matrix (left inset). We can use a $2\times2$ propagation matrix to model the propagation within a medium (right inset). $k$ is the wavenumber, $\eta$ is the medium impedance and $l$ is thickness of the medium.}
  \label{fig:interface}
\end{figure}

In this section, impedance boundary conditions (IBCs) and generalized sheet transition conditions (GSTCs) are reviewed. Then scattering coefficients of one-dimensional interfaces are defined, and related formulas derived from IBCs and GSTCs are given. The procedure starts by defining the boundary-value problem associated with object illusion problems. This includes state of the art wave-scattering modelling within confined complex environments. Conventional boundary conditions describe the relationship between field discontinuities across boundary and current sources distributed along boundary, which can be expressed as
\begin{equation}
  \label{eq:jse}
  \hat{n}\times(\Vec{H_2}-\Vec{H_1}) =\Vec{J_{se}} 
\end{equation}
\begin{equation}
  \label{eq:jme}
  (\Vec{E_2}-\Vec{E_1})\times\hat{n} =\Vec{J_{sm}} 
\end{equation}
where ($E_1$, $H_1$) and ($E_2$, $H_2$) represent the total fields at both sides of the interface, while $J_{se}$ and $J_{sm}$ are the surface electric and magnetic current densities along the boundary. The field jump across EM surfaces can be characterized by their effective sheet impedances, based on which IBCs as derived in \cite{13}. The electric sheet impedance $Z_e$ is defined as the ratio of average electric fields $E_{av}$ along the boundary to surface electric current density $J_{se}$, while the magnetic sheet impedance $Z_m$ is the ratio of $J_{sm}$ to average magnetic fields $H_{av}$. Consequently, the IBCs describe the relationship between average fields along surfaces and field discontinuities across surfaces
\begin{equation}
  \label{eq:Ze}
  \hat{n}\times(\Vec{H_2}-\Vec{H_1}) = Z_e^{-1}E_{av}
\end{equation}
\begin{equation}
  \label{eq:Zm}
  (\Vec{E_2}-\Vec{E_1})\times\hat{n} =Z_mH_{av}
\end{equation}
Note that all field components that appear in IBCs are tangential. With IBCs, field discontinuities across EM surfaces can be determined once their sheet impedances $Z_e$ and $Z_m$ are known. However, they are not characteristic parameters of EM surfaces and their values for different situations cannot be readily determined. The characteristic parameters of one-dimensional interfaces are effective susceptibilities, whose values are independent of applied fields. Based on susceptibilities, GSTCs are initially derived for metafilms \cite{14}.
\begin{equation}
  \label{eq:Xe}
  \hat{n}\times(\Vec{H_2}-\Vec{H_1}) = j\omega\epsilon_0\chi_eE_{av}-\hat{n}\times\nabla_t(\chi_mH_{av})
\end{equation}
\begin{equation}
  \label{eq:Xm}
  (\Vec{H_2}-\Vec{H_1})\times\hat{n} = j\omega\mu_0\chi_mH_{av}+\hat{n}\times\nabla_t(\chi_eE_{av})
\end{equation}
where one-dimensional electric $\chi_e$ and magnetic $\chi_m$ tensors represent the effective surface electric and magnetic susceptibilities respectively. We can calculate the transmission and reflection coefficients for this interface as \cite{15}
\begin{equation}
  \label{eq:tau}
  \tau_{M} =\frac{1-(k_0/2)^2\chi_{e}\chi_m}{1+(k_0/2)^2\chi_{e}\chi_m-jk_0/2(\chi_m-\chi_e)}
\end{equation}
\begin{equation}
  \label{eq:rho}
  \rho_{M} =\frac{jk_0/2(\chi_m+\chi_e)}{1+(k_0/2)^2\chi_{e}\chi_m-jk_0/2(\chi_m-\chi_e)}.
\end{equation}
Using the IBCs for the fields, we can relate the forward-backward fields on one side of the interface to those on the other side, expressing the relationship in terms of a $2\times2$ transfer matrix. Figure~\ref{fig:interface} shows both the interface and the associated shift matrices in terms of reflection $\rho$ and transmission coefficients \(\tau\).

\begin{figure}[htbp]
\vspace{-0.7cm}
  \centering
  \includegraphics[trim={6.5cm 4cm 7.4cm 3cm},clip,width=82mm]{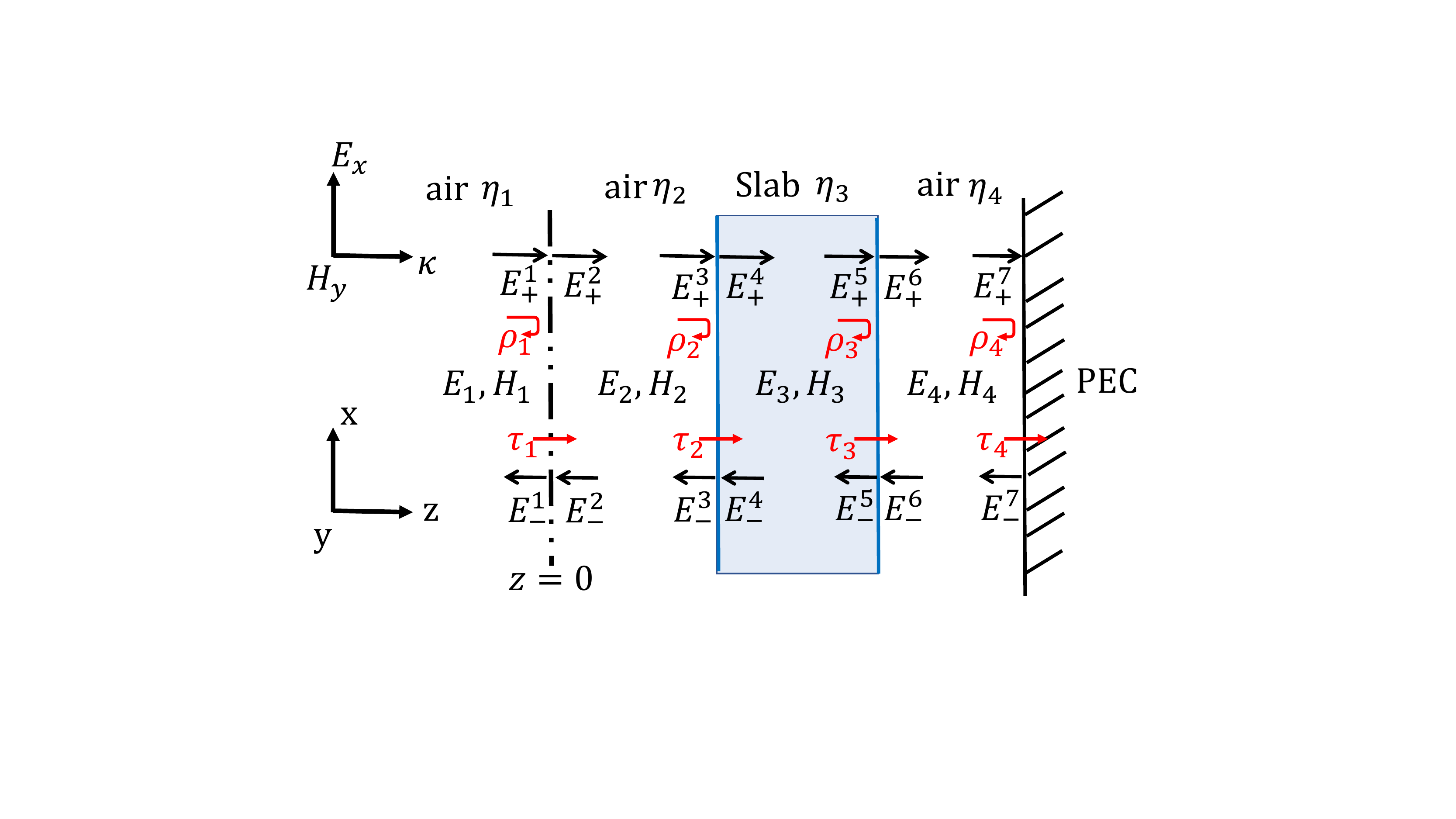}
  \vspace{-0.7cm}
  \caption{Simulation model.}
  \label{fig:cavity}
\end{figure}

If there are several interfaces, we can propagate our forward-backward fields from one interface to the next with the help of a $2\times2$ propagation matrix. The combination of a shift and a propagation matrix relating the fields across different interfaces will be referred to as a transfer or transition matrix. Multiple interface problems can be handled in a straightforward way with the help of the transfer matrix. Figure~\ref{fig:cavity} shows the simulation model for an arbitrary slab in the air that has a discontinuous interface on one side and a PEC wall on the others. 

\section{Results}
\begin{figure}[htbp]
\vspace{-0.25cm}
  \centering
  \includegraphics[trim={1cm 6.5cm 2cm 6.5cm},clip,width=41mm]{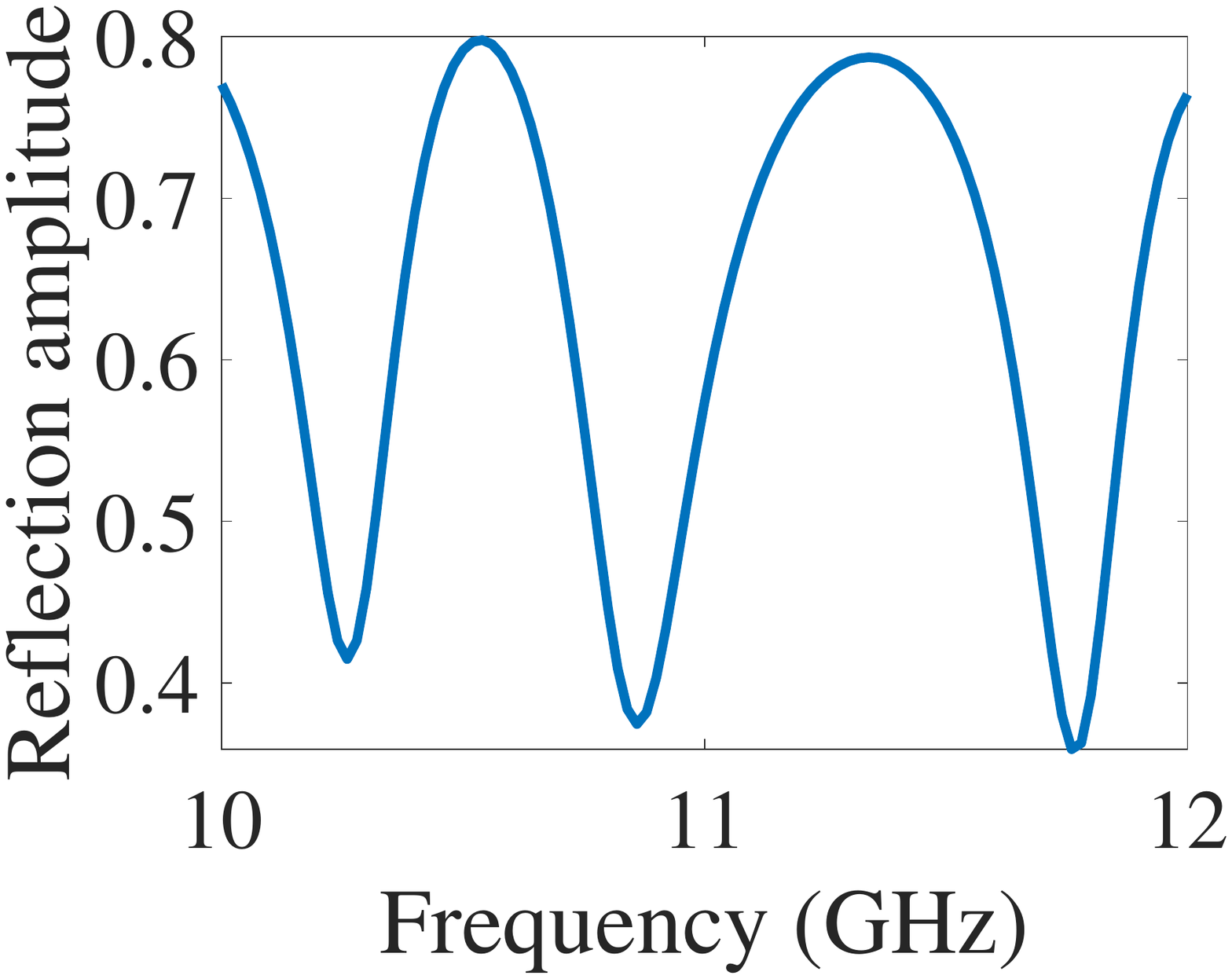}
    \includegraphics[trim={1cm 6.5cm 2cm 6.5cm},clip,width=41mm]{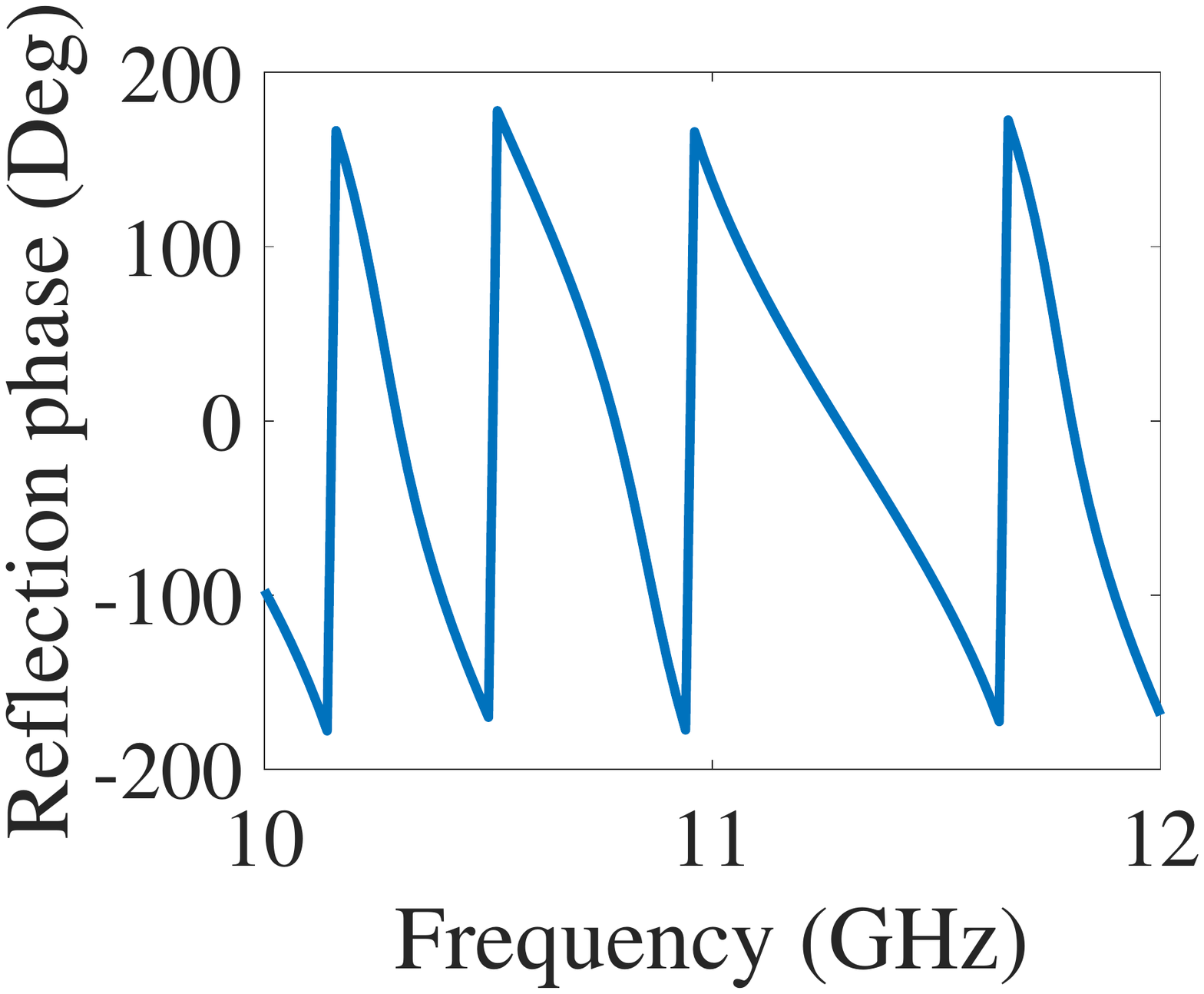}
  \caption{Reflection coefficients from the FR4 slab inside air and background is a PEC wall.}
  \label{fig:refslab}
\end{figure}

In the first case, we study the possibility of hiding the slab by placing a metasurface along the PEC wall. The slab is made of FR4 with permittivity $\epsilon_d=3.9-0.08i$ and thickness $D=50 mm$ also the air space from lateral sides is $D=100 mm$. A planar impinging wave encounters this slab and the wave partially transmits and then bounces back and forth inside the slab. The transmitted part hits the PEC and reflects back. The transfer matrix reads
\begin{multline}
  \label{eq:refmatrix}
  \begin{bmatrix}
E_+^1  \\
E_-^1  
\end{bmatrix}=\frac{1}{\tau_1\tau_2\tau_3}\times\\ \begin{bmatrix}
Z_1^{-1} & \rho_1Z_1 \\
\rho_1Z_1^{-1}  & Z_1
\end{bmatrix}
\begin{bmatrix}
Z_2^{-1} & \rho_2Z_2 \\
\rho_2Z_2^{-1}  & Z_2
\end{bmatrix}
\begin{bmatrix}
Z_3^{-1} & \rho_3Z_3 \\
\rho_3Z_3^{-1}  & Z_3
\end{bmatrix}
  \begin{bmatrix}
E_+^7  \\
E_-^7  
\end{bmatrix}
\end{multline}
where $Z_n=e^{-jk_nl_n}$ is obtained via the propagation matrix ($Z_1=Z_3$) and  $\rho_n$, $\tau_n$ are the local reflection and transmission coefficients defined as
\begin{equation}
  \label{eq:elementry}
  \rho_n=\frac{\eta_{n+1}-\eta_n}{\eta_{n+1}+\eta_n}, \tau_n=\frac{2\eta_{n+1}}{\eta_{n+1}+\eta_n}.
\end{equation}
We then calculate the total reflection at $z=0$ as $\Gamma=E_-^1/E_+^1$. Total reflection amplitude and phase are shown in Fig.!\ref{fig:refslab}. Resonances correspond to the stationary waves forming within a cavity-like structure and generated from the interference of forward and backward wave fields.

\begin{figure}[htbp]
\vspace{-0.25cm}
  \centering
  \includegraphics[trim={1cm 6.5cm 1cm 6.5cm},clip,width=41mm]{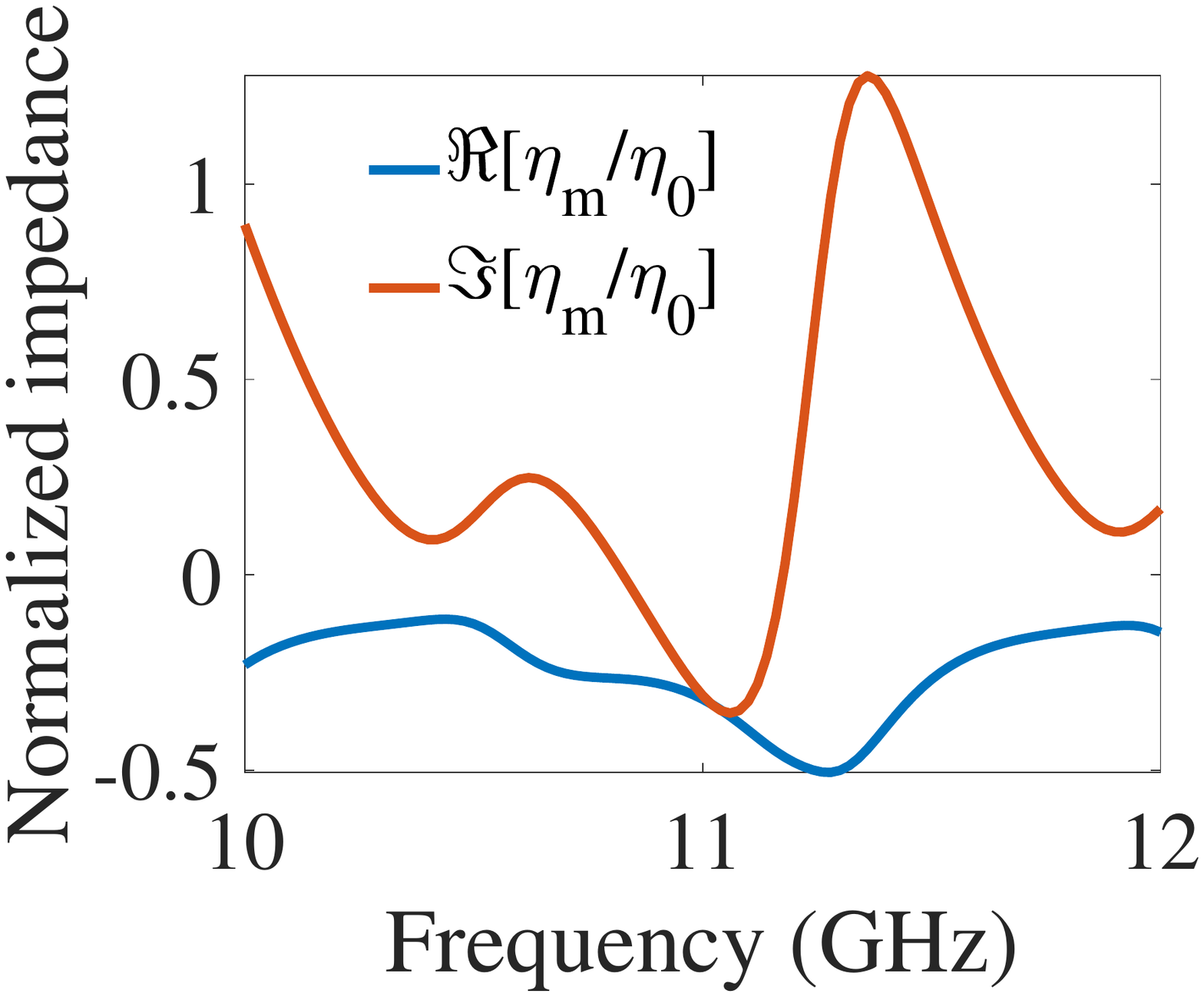}
    \includegraphics[trim={1cm 6.5cm 1cm 6.5cm},clip,width=41mm]{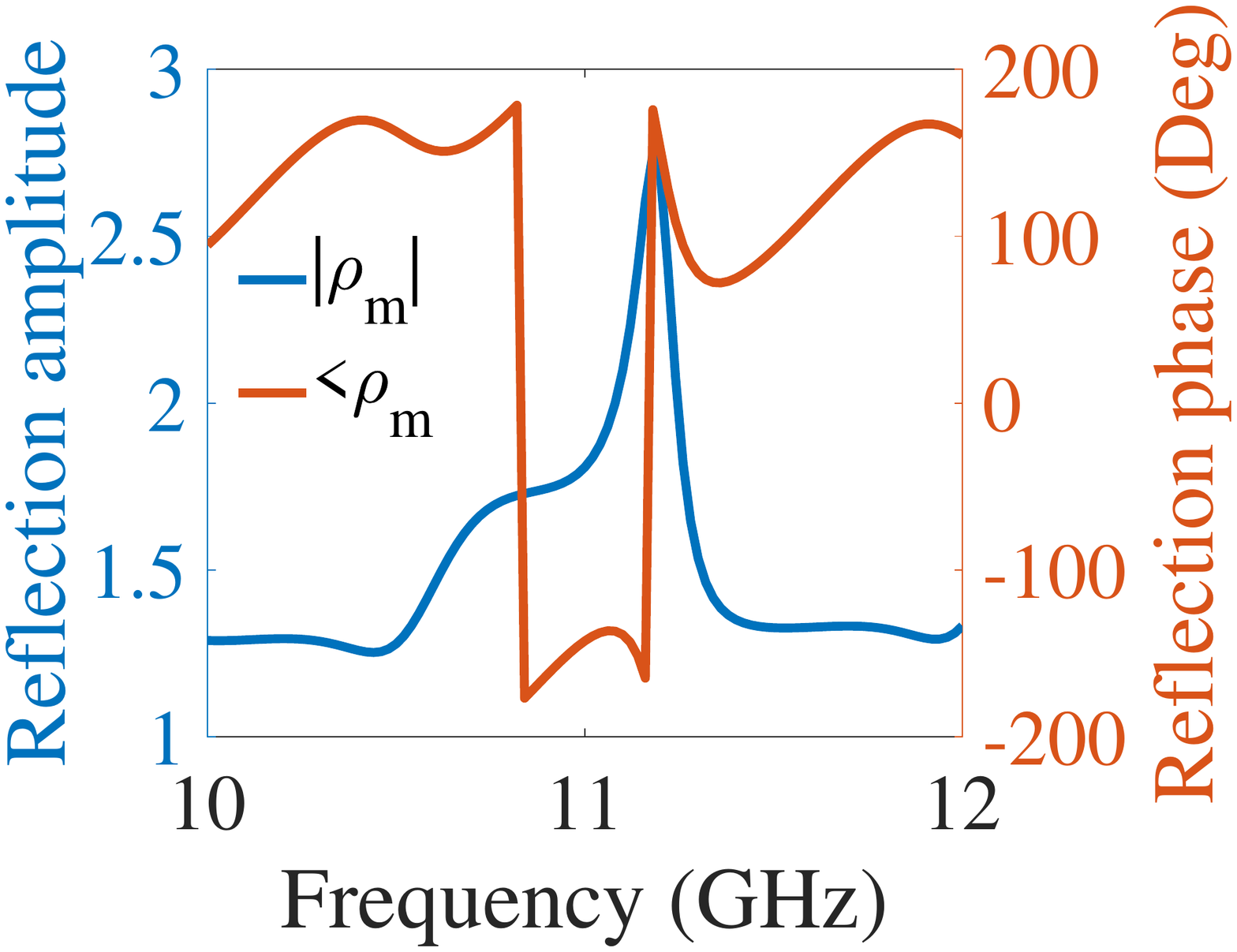}
  \caption{Surface impedance of the engineered metasurface and corresponding reflection coefficients.}
  \label{fig:msimp}
\end{figure}

Now, defining $\Gamma\equiv-Z_1^6$, we can find the required surface impedance of the metasurface to conceal the slab scattering. The essential surface impedance reads
\begin{equation}
  \label{eq:imp}
\eta_m=\eta_0\frac{A+B}{A-B}
\end{equation}
\begin{equation}
  \label{eq:A}
A=-Z_2Z_1^9(\rho_3+\rho_2Z_2^2)-Z_1^5Z_2(\rho_3\rho_2+Z_2^2)
\end{equation}
\begin{equation}
  \label{eq:B}
B=Z_2Z_1^3(\rho_2+\rho_3Z_2^2)+Z_1^7Z_2(1+\rho_3\rho_2Z_2^2)
\end{equation}
The surface impedance and the corresponding reflection coefficients are shown in Fig. \ref{fig:msimp}. The negative real part of the surface impedance implies that we need active elements to hide the slab and reflection amplitudes greater than one confirm this requirement.

\begin{figure}[htbp]
\vspace{-0.25cm}
  \centering
  \includegraphics[trim={1cm 6.5cm 1cm 6.5cm},clip,width=41mm]{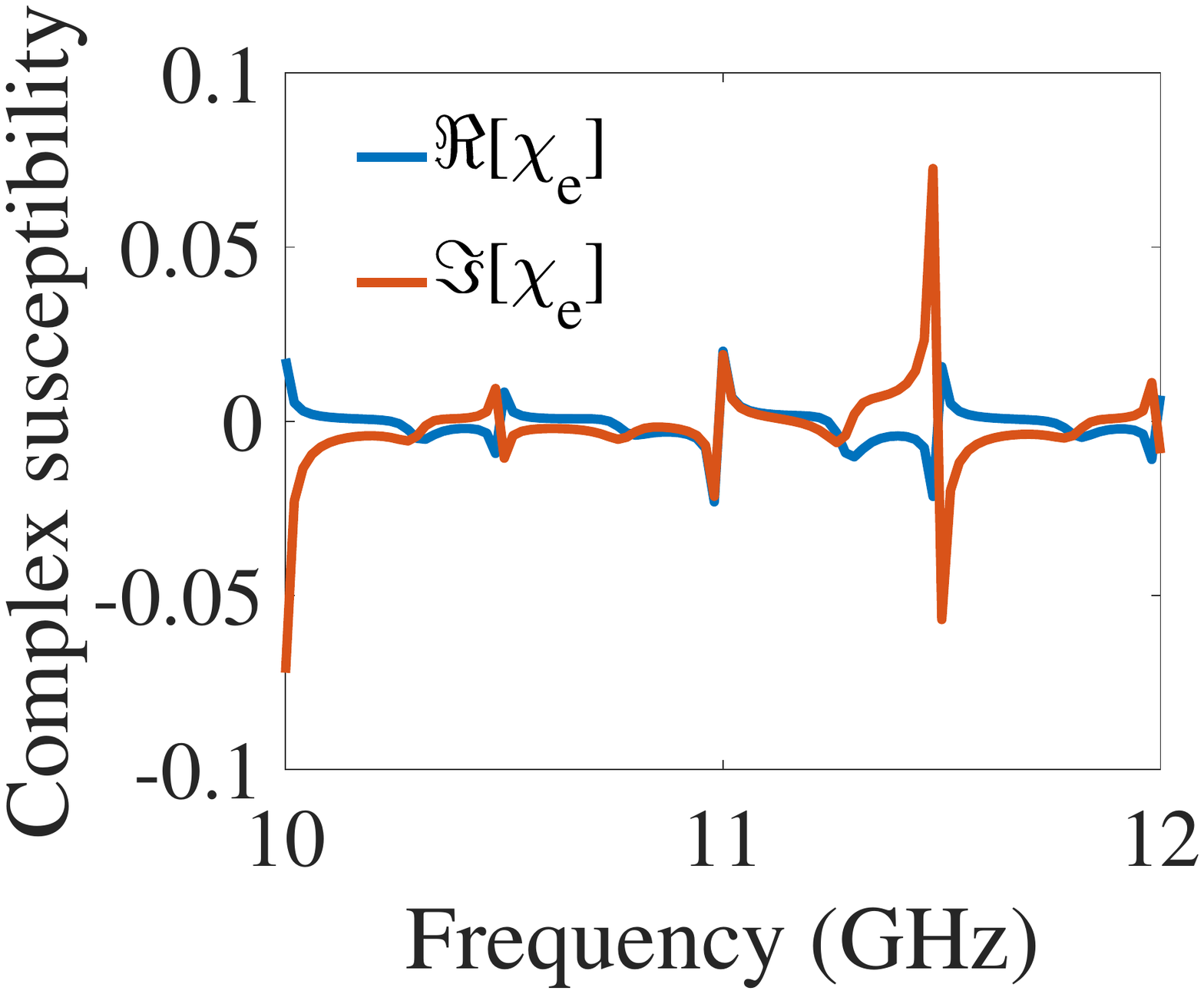}
    \includegraphics[trim={1cm 6.5cm 1cm 6.5cm},clip,width=41mm]{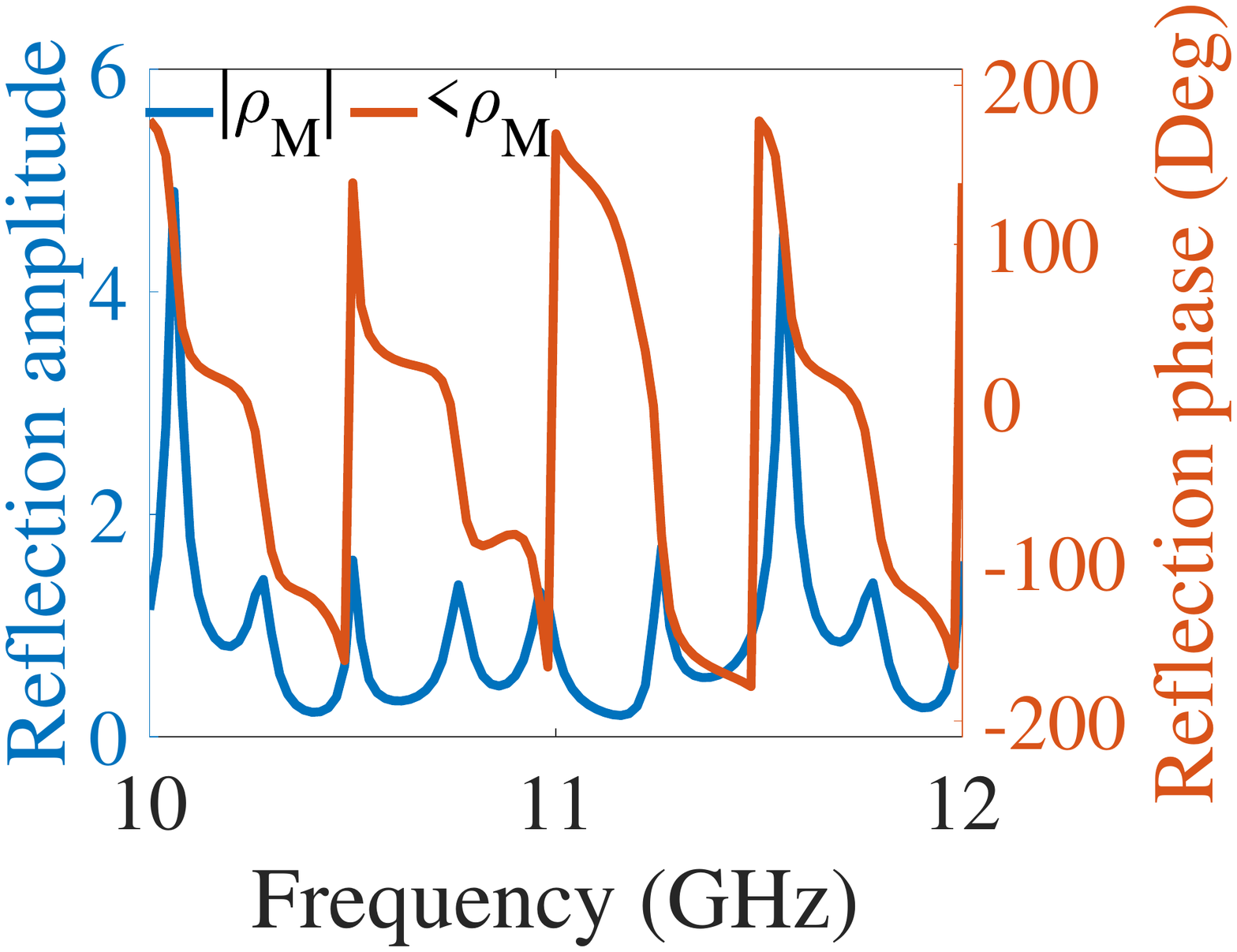}
  \caption{Surface susceptibility of the engineered metasurface and corresponding reflection coefficients.}
  \label{fig:mssus}
\end{figure}

We can also put a metasurface sheet before the slab and by applying GSTCs find the required susceptibilities to hide the slab from impinging wave. By replacing $\rho_1$ with $\rho_M$ in Eq.~(\ref{eq:rho}) and only considering the electric susceptibility ($\chi_m=0$)
\begin{equation}
  \label{eq:chi}
\chi_e=\frac{a}{j\frac{k_0}{2}(a+b)}
\end{equation}
where
\begin{multline}  
\label{eq:a}
a=Z_2(Z_1^2-\rho_3)+\rho_2Z_2^{-1}(\rho_3Z_1^2-1)-\\
Z_1^4(Z_2^{-1}+\rho_3\rho_2Z_2)+Z_1^6(\rho_3Z_2^{-1}+\rho_2Z_2)
\end{multline}
\begin{multline}  
\label{eq:b}
b=Z_2^{-1}(Z_1^{-2}-\rho_3)+\rho_2Z_2(\rho_3Z_1^{-2}-1)-\\
Z_1^8(Z_2+\rho_3\rho_2Z_2^{-1})+Z_1^6(\rho_2Z_2^{-1}+\rho_3Z_2).
\end{multline}
The real and imaginary part of the obtained electric susceptibility with corresponding reflection coefficients are shown in Fig.~\ref{fig:mssus}. Here, we also need active elements to achieve camouflage. While susceptibilities are not showing high values, their associated reflection amplitudes are high. This means that reflection coefficients of metasurfaces are more sensitive to susceptibility compared to surface impedance. 

\begin{figure}[htbp]
  \centering
  \includegraphics[trim={0cm 7.7cm 0cm 8.4cm},clip,width=82mm]{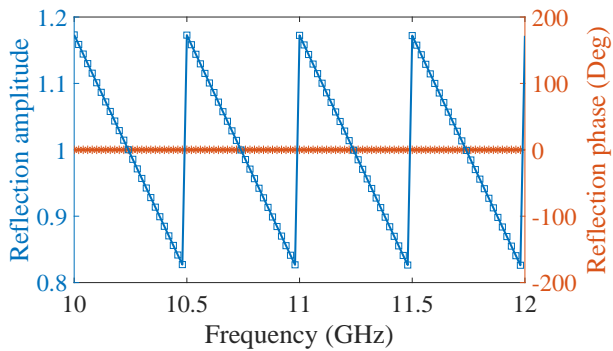}
  \caption{Reflection coefficients from the slab after placing the metasurfaces (solid lines). Reflection coefficients from the PEC without any slab in the medium (symbols).}
  \label{fig:cloak}
\end{figure}

Finally, we notice that modeling the metasurface either via a surface impedance $\eta_m$ over the PEC wall or am electric susceptibility $\chi_e$ before the slab, we can conceal the scattering from the slab as shown in Fig.~\ref{fig:cloak}. The amplitude of the reflection coefficient is unity without resonances and the reflection phase is as a PEC reflection, which shows that the dielectric slab is hidden via metasurface manipulation of scattered waves.

The proposed strategy is applicable to other illusion effects. This study shows that environmental engineering is a solid tool for object illusion without the need for direct access to the object itself. 

\section{Conclusion}
A new paradigm of smart environment for dynamic electromagnetic illusion is introduced.  We envision reconfigurable metasurfaces using active elements to change the interaction waves between object and environment and thus result in electromagnetic illusion. We proposed a proof of principle within a one-dimensional medium. In contrast to previous works, we avoided using any object coating. Also, the object is not necessarily in free space. An FR4 slab is hidden from radar sensors by placing the metasurface in the background or before the slab. This work paves the way for 2D or 3D metasurface designs with powerful illusion abilities.

\section{Acknowledgements}
This work has been supported by the EPSRC under Grant EP/V048937/1 and in part by the European Commission through the H2020 RISE-6G Project under Grant 101017011.

\end{document}